\documentclass[twocolumn,showpacs,preprintnumbers,amsmath,amssymb,superscriptaddress]{revtex4}

\usepackage{graphicx}
\usepackage{dcolumn}
\usepackage{bm}
\usepackage{epstopdf}
\usepackage{mathrsfs}
\usepackage{amssymb}
\usepackage{amsmath}

\def\bs{\boldsymbol}
\def\del{\partial}

\def\p{{\boldsymbol p}}
\def\pb{\bar {\boldsymbol p}}

\def\q{{\boldsymbol q}}

\def\k{{\boldsymbol k}}

\def\n{{\boldsymbol n}}
\def\x{{\boldsymbol x}}

\def\bkappa{{\boldsymbol \kappa}}
\def\bbkappa{\bar{\boldsymbol \kappa}}

\def\qqb{{q\bar q}}
\def\sM{\text{med}}

\newcommand{\beq}{\begin{eqnarray}}
\newcommand{\eeq}{\end{eqnarray}}
\newcommand{\be}{\begin{eqnarray*}}
\newcommand{\ee}{\end{eqnarray*}}

\begin{document}


\title{Jets in QCD media: from color coherence to decoherence}

\author{Yacine Mehtar-Tani}
\author{Carlos A. Salgado}%
\affiliation{%
Departamento de F\'isica de Part\'iculas and IGFAE,
Universidade de Santiago de Compostela \\
E-15782 Santiago de Compostela, 
Galicia-Spain
}%
\author{Konrad Tywoniuk}
\affiliation{%
Department of Theoretical Physics, 
Lund University. 
SE-223 62 Lund, 
Sweden
}%

\date{\today}

\begin{abstract}
We investigate soft gluon radiation off  a quark--antiquark antenna in both color singlet and octet configurations traversing a dense medium. We demonstrate that, in both cases,
 multiple scatterings lead to a gradual decoherence of the antenna radiation as a function of the medium density. In particular, in the limit of an completely opaque medium, total decoherence is obtained, i.e., the quark and the antiquark radiate as independent emitters in vacuum, thus losing memory of their origin. We comment on possible implications on intrajet observables in heavy-ion collisions.

\end{abstract}

\maketitle

The LHC has reported the first data on reconstructed jets in Pb+Pb collisions at $\sqrt{s} = 2.36$ TeV \cite{Aad:2010bu}. A significant imbalance in dijet transverse momentum is observed, implying a modification of the in-medium fragmentation process compared to the one in vacuum. 
Preliminary results from RHIC are also available \cite{Putschke:2008wn}.
These new experimental observables call for a well controlled theoretical description of jet formation in such conditions. The purpose of this letter is to address, from first principles,
intrajet coherence effects in the presence of a QCD medium.

In QCD, the sequential radiation process in vacuum is accompanied by soft and collinear singularities \cite{bas83,Dokshitzer:1978hw}. Furthermore, intrajet coherence effects give rise to strict angular ordering of subsequent emissions \cite{Mueller:1981ex,Ermolaev:1981cm}, thus severely reducing the phase space for soft gluons.

It is well known that the radiation process is intensified in the presence of a deconfined medium. However, until very recently, these modifications were only computed for the single-gluon radiation off a highly energetic quark or gluon in the soft and collinear limit \cite{bdmps,zakharov,gyu00,Wiedemann:2000za}. The resulting spectrum is both infrared and collinear safe and, moreover, does not embody intrajet coherence.

Owing to the non-divergent structure of the medium-induced radiation described above multiple emissions cannot be resummed similarly to the vacuum cascade. To date, the generalization to medium-induced multi-gluon radiation rely on {\it ad hoc} conjectures. In our recent paper \cite{MehtarTani:2010ma}, we have taken a first step towards resolving this disparity by calculating the coherent gluon spectrum off a quark--antiquark ($\qqb$) antenna at first order in the background field,
which turned out to be infrared divergent.

In the limit of small opening angle of the antenna, we found that the incoherent medium-induced spectrum is cancelled due to destructive interference \cite{MehtarTani:2010ma}. The resulting spectrum has an interesting structure which we have called {\it anti--angular ordering}. Namely, medium-induced gluon radiation off each of the emitters is forbidden inside the cone delimited by the antenna opening angle after averaging over the corresponding azimuthal angle. Moreover, it also exhibits a vacuum-like emission pattern outside.

In this work we extend our previous results in the soft sector aiming at describing gluon emissions in the presence of opaque media, rendering the physical picture more transparent. This demands the resummation of multiple interactions with the medium. The main result of this letter is that color coherence, essential for jets in vacuum, is gradually washed out with increasing medium density leading to total decoherence in the limit of an infinitely opaque medium. This conclusion holds for antennas in arbitrary color representation, i.e., originating from the splitting of a virtual photon or gluon, and also a quark.

As in \cite{MehtarTani:2010ma}, we set our calculation within the framework of classical Yang-Mills (CYM) equations, applicable for soft gluon radiation off an energetic charge \cite{meh07}. First, let us recall the amplitude of emitting a gluon with momentum $k\equiv(\omega,\vec k)$, given by
\beq
\label{eq:redform}
{\cal M}_\lambda ^{a}({\vec k})=\lim_{k^2\to 0 } -k^2A^{a}_\mu(k)  \epsilon^\mu_\lambda({\vec k}) \,,
\eeq
where $\epsilon^\mu_\lambda({\vec k})$ is the gluon polarization vector while $A^a_\mu$, the classical gauge field, is the solution of the CYM equations, $[D_\mu,F^{\mu\nu}] = J^\nu$, with $D_\mu\equiv \del_\mu-ig A_\mu$ and $F_{\mu\nu}\equiv \del_\mu A_\nu-\del_\nu A_\mu-ig[A_\mu,A_\nu]$. The covariantly conserved current, i.e., $[D_\mu,J^\mu]=0$,  describes the projectiles. Furthermore, we shall set our calculation in the light-cone gauge (LCG), $A^+\equiv (A^0+A^3)/\sqrt{2}=0$, where only the transverse polarization contribute to the cross-section, and $\sum_{\lambda} \epsilon^i_\lambda(\epsilon_\lambda^j)^\ast=\delta^{ij}$, where $i(j)=1,2$.  

Consider a $q \bar q$ pair with momenta $p\equiv(E, \vec p)$ and $\bar p\equiv( \bar E, \vec{\bar p})$, respectively, created in the splitting of a virtual photon or gluon moving in the $+z$ direction. In the absence of the medium, the classical eikonalized current that describes the pair created at time $t_0=0$ reads $J^\mu_{(0)}=J^\mu_{q\,(0)}+J^\mu_{\bar q \,(0)}$, where $ J^{\mu,a}_{q \,(0)} = g\frac{p^\mu}{E}\,\delta^{(3)}({\vec x}-\frac{\vec p}{E}t)\,\Theta(t) \, Q_{q}^a$. Here, $Q_q$ denotes the color charge vector of the quark (and, analogously, $Q_{\bar q}$ for the antiquark). The sum of quark and antiquark charges gives the color charge of the whole system, i.e., the charge of the parent projectile, $Q_q + Q_{\bar q} = Q_I$, where $I$ denote the color representation of the pair ($I\equiv \gamma^\ast,g^\ast$). 
In the case of a highly virtual $g^\ast\to\qqb$ splitting, the third component of the current, ensuring color conservation, is implicit and will not contribute in the forward region thanks to the gauge choice.

Concerning the singlet configuration, $\gamma^\ast\to\qqb$, the color charge of the antenna is vanishing, $Q_{\gamma^\ast} = 0$, thus $Q_q = - Q_{\bar q}$. In the octet case, on the other hand, it is given by $Q_{g^\ast}$, namely the color vector of the parent gluon.  Then, by taking the square of the total color charge, we get that $2\,Q_q\cdot Q_{\bar q} = C_A -2\,C_F$, since $Q_q^2 = C_F \equiv (N_c^2-1)/2N_c$ and $Q_g^2 = C_A\equiv N_c$. Other color configurations, e.g., $g^\ast\to gg$, can be considered in a similar fashion.

We use light-cone variables defined by $k=(k^+, k^-,\k)$, where $k^{\pm}\equiv (\omega\pm k^3)/\sqrt{2}$ and $\k\equiv(k^1,k^2)$, and similarly for any vector in what follows. At leading order in the coupling, $g$, the linearized  CYM equations yield, together with Eq.~(\ref{eq:redform}),
\beq
\label{eq:ampvac-soft}
{\cal M}_{(0)\lambda}^{a} (k) = -ig\left[\frac{ \bkappa\cdot {\bs \epsilon_\lambda}}{ x\ (p\cdot k)}\ Q_q^a  + \frac{ \bbkappa\cdot{\bs \epsilon_\lambda}}{\bar x \ (\bar p\cdot k)}\ Q_{\bar q}^a \right]  \,,
\eeq
where we have introduced the following transverse vectors $\kappa^i \equiv k^i - x\, p^i$ and $\bar \kappa^i \equiv k^i - \bar x\, \bar p^i$ ($i=1,2$), along with the momentum fractions 
$x\equiv k^+/p^+\approx \omega/E$ and $\bar x \equiv k^+/\bar p^+\approx \omega/\bar E$ (which are implicit in the rest of the paper). 
Summing over the gluon polarization vectors, it can be easily checked that the well known cross section for the color octet case reads \cite{Dokshitzer:1991wu}
\begin{align}
\label{eq:spectrum-vac1}
(2\pi)^2\,\omega\frac{dN_{g^\ast}^\text{vac}}{d^3k}= \frac{\alpha_s}{\omega^2}\left[ C_F\mathcal{R}_\text{coh} + C_A \mathcal{J}\right] \,,
\end{align}
where $\mathcal{R}_\text{coh} \equiv \mathcal{R}_q + \mathcal{R}_{\bar q} - 2 \mathcal{J}$ with $\mathcal{R}_q = 2/(n_q\cdot n)$, and analogously for the antiquark, and $\mathcal{J} = \bkappa\cdot \bbkappa/[\omega^2(n_q\cdot n)(n_{\bar q}\cdot n)]$, where $n^\mu_q = p^\mu/E$ and $n^\mu = k^\mu/\omega$. In the color singlet case, we are left with only the first term in Eq.~(\ref{eq:spectrum-vac1}), i.e., the one proportional to $C_F$.

Let us quickly recall some key features of Eq.~(\ref{eq:spectrum-vac1}). In the singlet case, the two collinear poles in $\mathcal{R}_\text{coh}$ can be split into two terms, $\mathcal{P}_q=\mathcal{R}_q - \mathcal{J}$ for the quark and analogously for the antiquark, which comprise the quark and the antiquark collinear divergences, respectively. Averaging $\mathcal{P}_q$ over the azimuthal angle leads to gluon emissions confined to a cone defined by the opening half-angle of the $\qqb$ pair, $\theta_{q\bar q}$, so that the spectrum reads
\beq
\label{eq:spectrum-vac2}
dN^{\text{vac}}_{q,\gamma^\ast}=\frac{\alpha_sC_F}{\pi} \frac{d\omega}{\omega}\frac{\sin\theta \ d \theta}{1-\cos\theta} \Theta(\cos\theta-\cos\theta_{q\bar q}),
\eeq
where $\theta$ is the angle between the quark and the emitted gluon. In the octet case, the additional term, coming with the adjoint color factor, $C_A$, gives rise to large-angle gluon emissions, $\langle \mathcal{J}\rangle_{\text{azimuth}}=2\, \Theta(\cos\theta_\qqb-\cos\theta)/(1-\cos \theta)$, which can be interpreted as the radiation off the total charge of the pair, which is that of the parent gluon \cite{Dokshitzer:1991wu}. Thus, the well-known feature of angular ordering applies to both color configurations.

We now turn to calculating medium interactions in the approximation of small opening angle of the pair, $\theta_{q\bar q}\ll1$, and for asymptotic energies of the quark and antiquark, respectively. Whereas in \cite{MehtarTani:2010ma} we restricted our calculation to first order in opacity, in the following we will include arbitrary number of rescatterings with the medium but considering only the soft sector. To do so, the pair field is treated as a perturbation around the strong static field $A_\sM$. In the asymptotic limit, the medium gauge field can be described by $A^-_\sM(x^+,\x)$, while $A_\sM^i=A^+_\sM=0$ \cite{meh07}. In Fourier space it reads
\beq
A^-_\sM(q) = 2\pi \, \delta(q^+)\int_{0}^{\infty} \!\!\! dx^+ ~{\cal A}_\sM(x^+,\q)~e^{i q^-\, x^+}.
\eeq
The continuity relation, given by $\del_\mu J^\mu = ig [A^-_\sM,J^+]$, can be  solved recursively with
\beq
J^{\mu}_{q(m)} = ig\frac{p^\mu}{p\cdot \del }~[A^-_\sM,J^+_{q(m-1)}] \;.
\eeq
Here, the subscript $m$ stands for the order of the expansion in the medium field. For $m>0$, in momentum space we get
\begin{multline}
\label{eq:current}
J^{\mu,a}_{q(m)}(k)= -(ig)^{m+1} \frac{p^\mu}{p\cdot k} \left[ \prod_{i=1}^{m}\int\frac{d^2\q_i}{(2\pi)^2} \int_0^{x^+_{i+1}} \!\!dx^+_i\right. \\
\times \left. e^{i \frac{\p\cdot\q_{i} }{p^+} x^+_i}T\cdot {\cal A}_\sM(x^+_i,\q_i) \right]^{ab} Q_q^b \ e^{i \frac{p\cdot k }{p^+} x^+_m},
\end{multline}
where $x^+_{m+1}=L$ is the total medium length and $T$ the generators of SU(3) in the adjoint representation. Summing over all possible interactions, $J_q^{\mu,a} = \sum_{m=0}^\infty J_{q(m)}^{\mu,a}$, yields
\begin{multline}
\label{eq:current-sol}
J^{\mu,a}_{q}(k) =- ig\frac{p^\mu}{p\cdot k }\\
\quad\times\left[ \delta^{ab}+\int_0^{L} dx^+\, e^{i\frac{p\cdot k}{p^+} x^+} \del^- U_p^{ab}(x^+,0)\right] \, Q_q^b\,,
\end{multline}
where the Wilson line along the trajectory of the quark in the adjoint representation, $U_p$, is given by
\beq
U_p(x^+,0)={\cal P}_+\exp\left[ig\int _0^{x^+} \!\!dz^+\, T\cdot A^-_\sM\left(z^+,\frac{\p}{p^+} z^+\right)\right] \,.
\eeq
In the strictly soft limit, i.e., $\omega\to 0$, Eq. (\ref{eq:current-sol}) reduces to 
\beq\label{eq:current-sol-soft}
J^{\mu,a}_{q}(k)& =&- ig\frac{p^\mu}{p\cdot k }U_p^{ab}(L,0)\, Q_q^b \,.
\eeq
The transverse component of the linear medium response reads \cite{meh07}
\beq
\label{eq:field1}
\square A^i=2ig\left[A_\sM^-,\del^+ A^i\right] -\frac{\del^i}{\del^+}J^++J^i \,.
\eeq
The first term in the r.h.s. of Eq. (\ref{eq:field1}) describes gluon rescattering with the medium, which screens the soft divergence rendering it infrared finite \cite{MehtarTani:2010ma}. The second and third terms correspond to gluon bremsstrahlung where only the quark rescatters and exhibits a soft divergence, see Eq.~(\ref{eq:current-sol-soft}). Keeping only the bremsstrahlung contribution, the amplitude for soft gluon emission off the quark and antiquark reads
\begin{multline}
\label{eq:amp-soft1}
{\cal M}_\lambda^{a}(k) = \\
-ig\left[\frac{\bkappa\cdot{\bs \epsilon}_\lambda}{x\ (p\cdot k) }U^{ab}_p(L,0)\ Q_q^b
 +\frac{\bbkappa\cdot{\bs \epsilon}_\lambda}{\bar x\ (\bar p\cdot k )}U^{ab}_{\bar p}(L,0)\ Q_{\bar q}^b\right] \,.
\end{multline}
This generalizes Eq.~(\ref{eq:ampvac-soft}) which we recover by putting $U=1$, i.e., in the absence of the medium.

Let us now discuss the color singlet antenna in medium. The spectrum in the soft limit is readily found from Eq.~(\ref{eq:amp-soft1}) to be
\begin{align}
\label{eq:finalspectrum}
(2\pi)^2\omega \frac{dN_{\gamma^\ast}^\text{tot}}{d^3k} &= \frac{ \alpha_s C_F}{\omega^2} \left(\mathcal{R}_\text{coh} + 2\Delta_\text{med}\,\mathcal{J}\right) \,,
\end{align}
where we have used that $Q_q^a Q_{\bar q}^b = \delta^{ab}/(N_c^2-1)\, Q_q\cdot Q_{\bar q}$.
The interaction with the medium is completely contained in $\Delta_\text{med}$, given by
\beq
\label{eq:Delta}
&&\Delta_{\text{med}}=1-\frac{1}{N_c^2-1}\langle {\bf Tr}\,U_p(L,0) U_{\bar p}^\dag(L,0) \rangle \,,
\eeq
which only affects the interference term, $\mathcal{J}$.
The brackets in Eq.~(\ref{eq:Delta}), $\langle ... \rangle$, stand for the medium expectation value, which we will discuss at length below. The color factor, $C_F$, appearing in Eq.~(\ref{eq:finalspectrum}), demonstrates that the emission takes place off the quark or the antiquark. Following the same decomposition as for the vacuum, leading to Eq.~(\ref{eq:spectrum-vac2}), the soft gluon spectrum off the quark in medium reads
\begin{widetext}
\beq
\label{eq:nqmed}
dN^\text{tot}_{q,\gamma^\ast}=\frac{\alpha_sC_F }{\pi}\frac{d\omega}{\omega}\frac{\sin\theta \ d \theta}{1-\cos\theta}\left[\Theta(\cos\theta-\cos\theta_{q\bar q})-\Delta_{\text{med}}\,\Theta(\cos\theta_{q\bar q}-\cos\theta)\right] \;.
\eeq
\end{widetext}
Equation~(\ref{eq:nqmed}) is a direct generalization of our previous result in the soft limit \cite{MehtarTani:2010ma} to multiple interactions. It has a simple form and offers an intuitive physical picture. Interestingly enough, the information about the medium is fully contained in a multiplicative factor, $\Delta_\text{med}$, while the functional shape is vacuum-like. In the dilute limit, $\Delta_\text{med}\to0$, we recover the pure vacuum spectrum, $dN_{q,\gamma^\ast}^\text{tot}\to dN_{q,\gamma^\ast}^\text{vac}$. With increasing density, the decoherence rate is controlled by the parameter $\Delta_\text{med}$. In the limit of a completely opaque system, $\Delta_\text{med}$ is bounded by unitarity so that $\Delta_\text{med} \to 1$. Then the soft emission in the presence of a medium reduces to independent radiation off the quark and antiquark, as if they were radiating in the vacuum. This is what we call {\it total decoherence} of the spectrum
\beq
\label{eq:spectrumdecoh}
dN^\text{tot}_{q,\gamma^\ast}\Big|_\text{opaque} = \frac{\alpha_sC_F }{\pi}\frac{d\omega}{\omega}\frac{\sin\theta \ d \theta}{1-\cos\theta} \,.
\eeq
In other words, the strict angular ordering condition is entirely removed. Thus, $\Delta_\text{med}$ appears as an order parameter controlling the transition between a coherent and decoherent situation.

\begin{figure}[t!]
\centering
\includegraphics[width=0.43\textwidth]{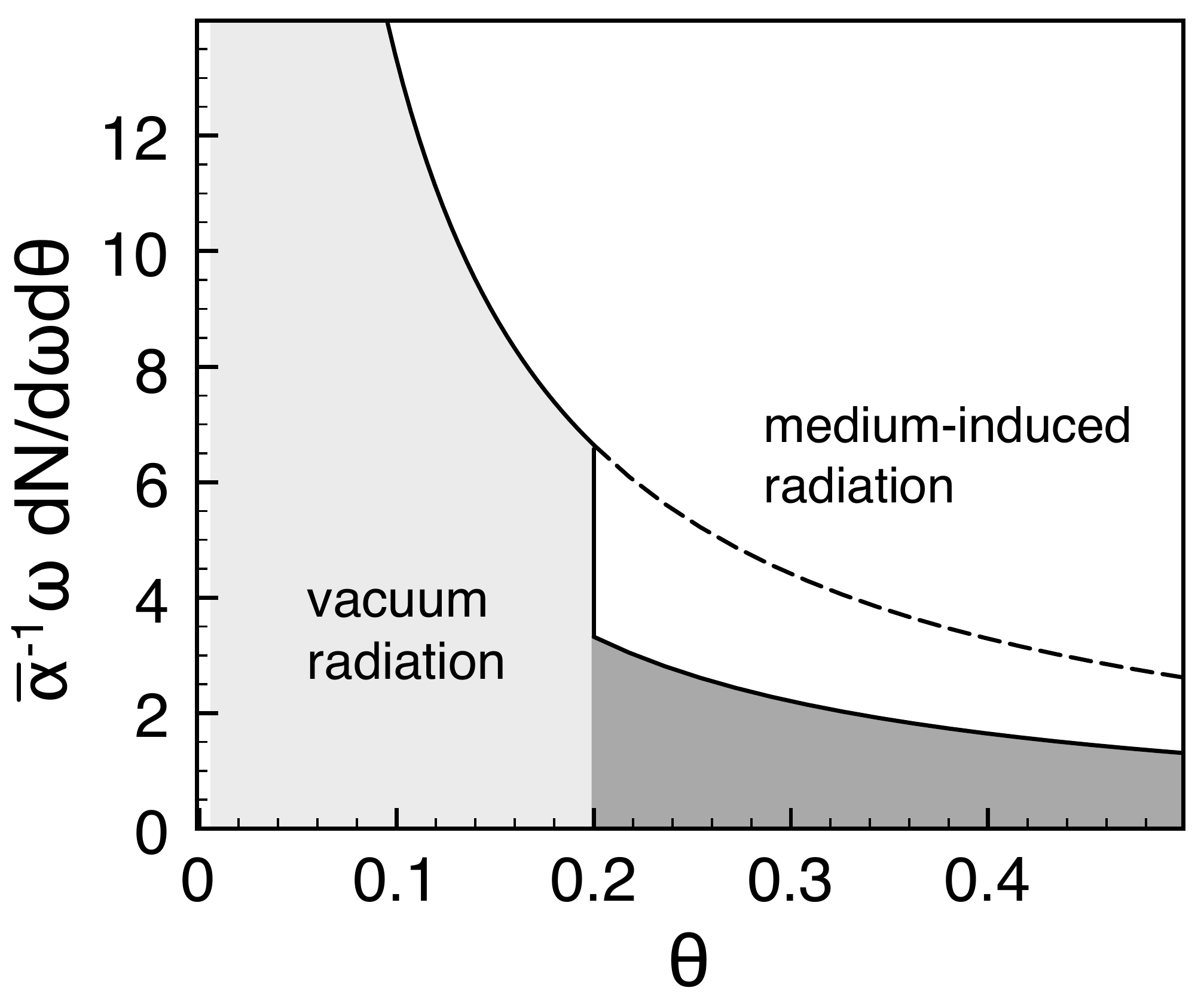}
\caption{The soft gluon emission spectrum off a energetic quark in the presence of a medium for a $\qqb$ pair with opening angle $\theta_\qqb = 0.2$ and $\Delta_\text{med} = 0.5$ (solid line). Here $\bar \alpha \equiv \alpha_s C_F/\pi$. Vacuum radiation is confined within $\theta < \theta_\qqb$, while the medium-induced radiation is radiated at $\theta > \theta_\qqb$. The limit of opaque medium, given by $\Delta_\text{med} = 1$, is marked by the dashed line.}
\label{fig:AngSpec}
\end{figure}
The general features of the spectrum interpolating between the dense and dilute medium limits are illustrated in Fig.~\ref{fig:AngSpec}, where we plot the angular spectrum of soft gluon emission off the quark for a $\qqb$ antenna with opening angle $\theta_\qqb = 0.2$. For $\theta < \theta_\qqb$, the spectrum is completely given by vacuum emissions, falling off as $1/\theta$. At $\theta = \theta_\qqb$ the medium-induced radiation takes over, controlled by the medium parameter $\Delta_\text{med}$. The limit of dense media is delineated by the dashed curve in Fig.~\ref{fig:AngSpec}. In this case, $\Delta_\text{med} =1$ and the total spectrum drops monotonously like $1/\theta$ without any discontinuity.

So far we have considered the generic behavior of the soft gluon spectrum without going into the details of how the medium is modelled. We proceed with a specific implementation, aiming at separating the medium properties from the dynamics of the rescattering of the $\qqb$-dipole. In particular, assuming that the medium is made out of uncorrelated static scattering centers, we can treat the background field, $A_\sM$, as  a Gaussian white noise
\begin{multline}
\langle {\cal A}^a_\sM(x^+,\q) {\cal A}^{\ast b}_\sM(x'^+,\q')\rangle\equiv \\
       \delta^{ab}\,n(x^+) \,\delta(x^+-x'^+)(2\pi)^2 \,\delta^{(2)}(\q-\q'){\cal V}^2(\q)\,,
\end{multline}
where ${\cal V}(\q)$ is usually chosen to be a screened Coulomb potential and $n(x^+)$ is the 3-dimensional density of scattering centers.  In this case, we easily find that
\begin{multline}
\label{eq:dipolefull}
\frac{1}{N_c^2-1} \langle {\bf Tr}\,U_q(L,0)U_{\bar q}^\dag(L,0) \rangle=\\ \exp\left[-\! \int_0^{L}\!\!\!\!dx^+\hat q(x^+)\ \sigma(|\delta \n| \, x^+)\right],
\end{multline}
where
\beq
\sigma(|\delta \n|\, x^+) = \int \!\! \frac{d^2\q}{(2\pi)^2} {\cal V}^2(\q)(1-\cos(\delta\n\cdot\q \,x^+)) \,,
\eeq
and we have defined the transport coefficient $\hat q(x^+) \equiv \alpha_s C_A \,n(x^+)$. Notice, that this definition differs slightly from the one in \cite{sal03}.
Above, we have also introduced the transverse vector $\delta \n = \p/p^+ - \pb/\bar p^+$. Expanding Eq.~(\ref{eq:dipolefull}) to first order in the medium field, we recover our previous result in the soft limit \cite{MehtarTani:2010ma}. Furthermore, going beyond the leading-log approximation by including the exponent under the integral in Eq.~(\ref{eq:current-sol}), we notice that the spectrum is exponentially suppressed for $\omega \gtrsim \omega_\qqb$, where $\omega_\qqb = 1/(\theta_\qqb^2 L)$ \cite{MehtarTani:2010ma}.

The generalization of the above results to the octet case follows closely the discussion above and reads
\begin{multline}
\label{eq:specoctet}
(2\pi)^2\,\omega\frac{dN_{g^\ast}^\text{tot}}{d^3k} = \frac{\alpha_s}{\omega^2}\Big[ C_F\left(\mathcal{R}_\text{coh} + 2\Delta_\text{med}\,  \mathcal{J} \right) \\  + C_A(1-\Delta_\text{med})\, \mathcal{J} \Big] \;.
\end{multline}
The first term in Eq.~(\ref{eq:specoctet}), coming with the intensity $C_F$, corresponds to the emission off the quark and the antiquark while the second term to the emission by the total charge of the pair, i.e., the parent gluon. Note that, in the totally opaque medium, the latter contribution vanishes, decorrelating the quarks from their parent. The former term, as in singlet case, reduces in this limit to the superposition of independent vacuum spectra off the quark and the antiquark, respectively. This fact implies a sort of memory loss effect in the medium, so that
\beq
dN_{g^\ast}^\text{tot}\Big|_\text{opaque} = dN_{\gamma^\ast}^\text{tot}\Big|_\text{opaque} \,,
\eeq
i.e., the antenna radiation is independent of the total color charge.

The simplicity of the above results in the limit of opaque media are striking. We expect these effects to be crucial for jet observables in heavy-ion collisions. A decoherence of jet fragmentation implies a strong softening of the intrajet distribution, as well as an increase in multiplicity. These general features, which also will lead to an angular broadening of the jet, are in qualitative agreement with recent experimental data \cite{Aad:2010bu}.
We note that different attempts to implement color decoherence phenomenologically in Monte-Carlo codes were considered in \cite{Leonidov:2010he}.

In summary, we have computed the antenna spectrum at leading-logarithmic accuracy for both color singlet and octet configuration interpolating between the vacuum and the completely opaque medium, thus demonstrating the gradual onset of decoherence.
In particular, our full numerical results at first order in opacity \cite{MehtarTani:2010ma} shows an intricate cancellation of the medium-induced incoherent part of the spectrum, resulting in a vacuum-like emission pattern, cf. Eq.~(\ref{eq:nqmed}), up to high gluon energies.
We expect this conclusion also to hold for multiple scattering considered here. 

\begin{acknowledgments}
The authors would like to thank N. Armesto, Yu. L. Dokshitzer and A. H. Mueller for helpful discussions. 
This work is supported by Ministerio de Ciencia e Innovaci\'on of Spain; by Xunta de Galicia (Conseller\'{\i}a de Educaci\'on and Conseller\'\i a de Innovaci\'on e Industria -- Programa Incite); by the Spanish Consolider-Ingenio 2010 Programme CPAN; and by by the European Commission. CAS is a Ram\'on y Cajal researcher.

\end{acknowledgments}

\begin{thebibliography}{99}

\bibitem{Aad:2010bu}
  G.~Aad {\it et al.} [Atlas Coll.],
  [arXiv:1011.6182 [hep-ex]];
    S.~Chatrchyan {\it et al.} [CMS Coll.],
  [arXiv:1102.1957 [nucl-ex]].
  
\bibitem{Putschke:2008wn}
  J.~Putschke
   [STAR Coll.],
  Eur.\ Phys.\ J.\  C {\bf 61}, 629 (2009); 
%
  S.~Salur  [STAR Coll.],
  {\it ibid.} {\bf 61}, 761 (2009);
  Y.~S.~Lai  [PHENIX Coll.],
  Nucl.\ Phys.\  A {\bf 830} (2009) 251C.

\bibitem{bas83}
A. Bassetto, M. Ciafaloni and G. Marchesini,  Phys. Rept.\ {\bf 100}, 201 (1983);
A. Bassetto, M. Ciafaloni, G. Marchesini and A.~H. Mueller, Nucl. Phys.\ {\bf B207}, 189 (1982).

\bibitem{Dokshitzer:1978hw}
  Yu.~L.~Dokshitzer, D.~Diakonov, S.~I.~Troian,
  Phys.\ Rept.\  {\bf 58 } (1980)  269-395.

\bibitem{Mueller:1981ex}
  A.~H.~Mueller,
  Phys.\ Lett.\  B {\bf 104}, 161 (1981).

\bibitem{Ermolaev:1981cm}
  B.~I.~Ermolaev and V.~S.~Fadin,
  JETP Lett.\  {\bf 33}, 269 (1981)
  [Pisma Zh.\ Eksp.\ Teor.\ Fiz.\  {\bf 33}, 285 (1981)].

\bibitem{bdmps}
R. Baier, 
Yu. L. Dokshitzer, A. H. Mueller, S. Peign\'e and D. Schiff, 
Nucl. Phys.\ {\bf B483}, 291 (1997); {\bf 484}, 265 (1997).

\bibitem{zakharov}
B. G. Zakharov, JETP Lett. \ {\bf 63}, 952 (1996); {\bf 65}, 615 (1997).

\bibitem{gyu00}
{M. Gyulassy, P. Levai and  I. Vitev},
Phys. Rev. Lett. {\bf 85}, 5535 (2000);
Nucl.\ Phys.\ {\bf B594}, 371 (2001).

\bibitem{Wiedemann:2000za}
  U.~A.~Wiedemann,
  Nucl.\ Phys.\ {\bf B588}, 303 (2000)

\bibitem{MehtarTani:2010ma}
  Y.~Mehtar-Tani, C.~A.~Salgado, K.~Tywoniuk,
  [arXiv:1009.2965 [hep-ph]].
  
\bibitem{meh07}
{Y. Mehtar-Tani},
Phys. Rev. C {\bf 75}, 034908 (2007). 

\bibitem{Dokshitzer:1991wu}
  Yu.~L.~Dokshitzer, V.~A.~Khoze, A.~H.~Mueller and S.~I.~Troyan,
{\it  Gif-sur-Yvette, France: Ed. Frontieres (1991) 274 p. (Basics of)}

\bibitem{sal03}
{C. A. Salgado and U. A. Wiedemann}, 
 Phys. Rev. D {\bf  68}, 014008 (2003). 
 
\bibitem{Leonidov:2010he}  
  K.~Zapp {\it et al.},
  Eur.\ Phys.\ J.\  {\bf C60}, 617-632 (2009);
    A.~Leonidov, V.~Nechitailo,
  [arXiv:1006.0366 [nucl-th]].

  

\end{thebibliography}

\end{document}